\documentclass[noshowpacs,preprintnumbers,amsmath,amssymb]{revtex4}

\usepackage{graphicx}
\usepackage{bm}
\usepackage{dcolumn}
\usepackage{color}
\usepackage{pst-plot}

\newcommand{\be}{\begin{equation}}
\newcommand{\ee}{\end{equation}}
\newcommand{\bea}{\begin{eqnarray}}
\newcommand{\eea}{\end{eqnarray}}
\newcommand{\non}{\nonumber}
\begin{document}
\title{Amplifying the Hawking signal in BECs}
\author{Roberto Balbinot}
\email{balbinot@bo.infn.it}
\affiliation{Dipartimento di Fisica dell'Universit\`a di Bologna and INFN sezione di Bologna, Via Irnerio 46, 40126 Bologna, Italy;\\  Museo Storico della Fisica e Centro Studi e Ricerche 'Enrico Fermi', Piazza del 
Viminale 1, 00184 Roma, Italy}
\author{Alessandro~Fabbri}
\email{afabbri@ific.uv.es}
\affiliation{Museo Storico della Fisica e Centro Studi e Ricerche 'Enrico Fermi', Piazza del 
Viminale 1, 00184 Roma, Italy;\\  Dipartimento di Fisica dell'Universit\`a di Bologna, Via Irnerio 46, 40126 Bologna, Italy; \\ Departamento de F\'isica Te\'orica and IFIC, Universidad de Valencia-CSIC, C. Dr. Moliner 50, 46100 Burjassot, Spain}
\date{\today}

\begin{abstract} 
We consider simple models of Bose-Einstein condensates to study analog pair-creation effects, namely the Hawking effect from acoustic black holes and the dynamical Casimir effect in rapidly time-dependent backgrounds. We also focus on a proposal by Cornell to amplify the Hawking signal in density-density correlators by reducing the atoms' interactions shortly before measurements are made.
\end{abstract}
\maketitle
\section{Introduction}
Analogue models in condensed matter systems are nowadays an active field of investigation, not only on the theoretical side but, more important, also on the experimental one.
The underlying idea is to reproduce in a condensed matter context peculiar and interesting quantum effects predicted by Quantum Field Theory in curved space, whose experimental verification in the gravitational context appears at the moment by far out of reach.

Many efforts are devoted to find the most famous of these effects, namely the thermal emission by black holes predicted by Hawking in 1974 \cite{Hawking:1974sw}.
Among the condensed matter systems under examination, Bose-Einstein condensates appear as the most promising setting to achieve this goal \cite{tre, blv}. 
The major problem one has to face experimentally is the correct identification of the signal corresponding to the analogue of Hawking radiation, namely a thermal emission of phonons
as a consequence of a sonic horizon formation, since it can be covered by other competing effects, like large thermal fluctuations.

A major breakthrough to overcome this problem came in 2008, when it was predicted that, as a consequence of being Hawking radiation a genuine pair creation process, a characteristic peak in the density correlation function of the condensate should appear for points situated on opposite sides with respect to the horizon \cite{Balbinot:2007de}.
This is the ``smoking gun'' of the Hawking effect. Soon after this proposal, Eric Cornell at the first meeting on ``experimental Hawking radiation'' held in Valencia in 2009, suggested that one can amplify this characteristic signal by reducing
the interaction coupling among the atoms of the BEC shortly before measuring the density correlations \cite{cornell}.

Here we review in a simple pedagogical way, using toy models, how the analogous of Hawking radiation occurs in a supersonic flowing BEC and how the corresponding characteristic peak in the correlation function can be amplified according to Cornell's suggestion. It should be stressed that nowadays correlation functions measurements are becoming the basic experimental tool to investigate Hawking-like radiation in condensed matter systems.

\section{BECs, the gravitational analogy and Hawking radiation}
\noindent  A Bose gas in the dilute gas approximation is described by a field operator $\hat \Psi$ with equal-time commutator (see for example \cite{ps})
\begin{equation}\label{etc}
[\hat \Psi (t,\vec x), \hat \Psi^{\dagger}(t,\vec x')]=\delta^3(\vec x- \vec x')
\end{equation}
satisfying the time-dependent Schr\"odinger equation 
\begin{equation}
i\hbar \partial_t \hat \Psi = \left(-\frac{\hbar^2}{2m}\vec \nabla^2 + V_{ext} + g\hat \Psi^{\dagger}\hat \Psi\right)\hat \Psi\ ,
\end{equation}
where $m$ is the mass of the atoms, $V_{ext}$ the external potential and $g$ the nonlinear atom-atom interaction  coupling constant. 
At sufficiently low temperatures a large fraction of the atoms condenses into a common ground state which is described, in the mean field approach, by a $c$-number field $\Psi_0(t,\vec x)$. \par  \noindent To consider linear fluctuations around this classical macroscopic condensate, one writes the bosonic field operator $\hat \Psi$ as 
\begin{equation}\label{mfexp} \hat \Psi \sim \Psi_0 (1 + \hat \phi)\ ,\end{equation}
where $\hat\phi$ is a small (quantum) perturbation.
 $\Psi_0$ and $\hat\phi$ satisfy, respectively, Gross-Pitaevski 
\begin{equation}\label{gp}
i\hbar\partial_t \Psi_0 = \left(-\frac{\hbar^2}{2m}\vec \nabla^2 + V_{ext} + g n_0 \right)\Psi_0\ 
\end{equation}
(where $n_0=|\Psi_0|^2$ is the number density) and Bogoliubov-de Gennes equations
\begin{equation}\label{bdg}
i\hbar  \partial_t   \hat \phi= - \left( \frac{\hbar^2}{2m}\vec \nabla^2 + \frac{\hbar^2}{m}\frac{\vec \nabla \Psi_0 }{\Psi_0} \vec \nabla\right)\hat\phi +mc^2 (\hat\phi + \hat\phi^{\dagger})\ ,
\end{equation}
with  $c=\sqrt{\frac{gn_0}{m}}$ is the speed of sound.

Contact with the gravitational analogy (see for example \cite{blv}) is achieved in the (long wavelength) hydrodynamic approximation, more easily realised by considering  the density-phase representation for the Bose operator $\hat\Psi=\sqrt{\hat n}e^{i\hat\theta}$ and the splitting $\hat n=n_0+\hat n_1,\ \hat\theta=\theta_0+\hat\theta_1$ in which $\hat n_1,\ \hat \theta_1$ represent the linear (quantum) density and phase fluctuations respectively. In terms of $\hat\phi$ and $\hat\phi^{\dagger}$ we have \begin{equation}\label{def} \hat n_1=n_0(\hat\phi+\hat\phi^{\dagger}),\ \hat \theta_1=-\frac{i}{2}(\hat\phi-\phi^{\dagger})\ .\end{equation}
Provided the condensate density $n_0$ and velocity $\vec v_0=\hbar \nabla \theta_0 /m$ vary on length scales much bigger than the healing length $\xi=\hbar/mc$ (the fundamental length scale of the condensate), the BdG equation reduces to the continuity and Euler equations for $\hat n_1$ and $ \hat \theta_1$ and these can be combined to give a second order differential equation for $\hat\theta_1$ which is mathematically equivalent to a Klein-Gordon (KG) equation \begin{equation}\Box\hat \theta_1=0\ , \end{equation} where $\Box$
is the covariant KG operator from the acoustic metric
\begin{equation}\label{acme}
ds^2=\frac{n_0}{mc}[-(c^2-\vec v_0^2)dt^2-2\vec v_0dtd\vec x +d\vec x^2]\ .
\end{equation}
For a flow which presents a transition from a subsonic ($|v_0|<c$) flow to a supersonic one 
($|\vec v_0|>c$ in some region) the metric (\ref{acme}) describes an acoustic black hole, with horizon located at the surface where $|v_0|=c$. 
The same analysis performed by Hawking in the gravitational case can be repeated step by step, leading to the prediction \cite{unruh} that acoustic black holes will emit a thermal flux of phonons at the temperature
\begin{equation} T_H=\frac{\kappa}{2\pi} \end{equation} where $\kappa=\frac{1}{2c}\frac{d(c^2-\vec v^2)}{dn}|_{hor},$ with $n$ the normal to the horizon, is the horizon's surface gravity.

\section{The model}

To simplify the mathematics involved in the process, 
we shall consider  a 1D  configuration \footnote{In 1D one should more correctly speak of quasi-condensation \cite{quattro}.} in which $n_0$ and $v_0$ are constant, and where the only nontrivial quantity is the speed of sound $c$. As explained in \cite{fnum}, this can be achieved by varying the coupling constant $g$ (and therefore $c$) and the external potential but keeping the sum $gn_0+ V_{ext}$ constant. In this way, the plane-wave function $\Psi_0=\sqrt{n_0}e^{ik_0x-iw_0t}$, where $v_0=\frac{\hbar k_0}{m}$ is the condensate velocity and $\hbar w_0=\hbar^2k_0^2/2m+V_{ext}+gn_0$, where $\hbar w_0$ is the chemical potential of the gas, is a solution of (\ref{gp}) everywhere. Note that such a stationary configuration is difficult to reach experimentally, nevertheless it gives results similar to those obtained by more realistic configurations \cite{cinque}. 

The fluctuation operator $\hat \phi$ is expanded in the usual form in terms of positive and negative  norm modes as
\begin{equation}\label{frep}
\hat\phi (t,x) =\sum_j \left[ \hat a_j \phi_j (t,x) + \hat a_j^{\dagger} \varphi_j^*(t,x) \right]\ ,\end{equation}
where $\hat a_j$ and $\hat a_j^{\dagger}$ quasi particle's annihilation and creation operators.
From (\ref{bdg}) and its hermitean conjugate, we see that the modes $\phi_j (t,x)$ and
$\varphi_j(t,x)$ satisfy the coupled differential equations
\bea\non
\left[ i(\partial_t + v_0\partial_x) + \frac{\xi c}{2} \partial_x^2 -\frac{c}{\xi} \right] \phi_j &=& \frac{c}{\xi}
\varphi_j\ , \\
\left[ -i (\partial_t + v_0\partial_x) + \frac{\xi c}{2}\partial_x^2 - \frac{c}{\xi}\right] \varphi_j &=&\frac{c}{\xi} \phi_j  \ .\label{cde}
\eea
The normalizations are fixed, via integration of the equal-time commutator obtained from (\ref{etc}), namely
\begin{equation}\label{etcd} [\hat \phi (t,x), \hat\phi^{\dagger}(t,x')]=\frac{1}{n_0}\delta(x-x')\ ,\end{equation}  by
\begin{equation}
\label{nor}
\int dx [\phi_j\phi_{j'}^* - \varphi_j^*\varphi_{j'}]=\frac{\delta_{jj'}}{ n_0}\ .\end{equation}

In order to get simple analytical expressions, in the following we shall consider simple models with step-like discontinuities in the speed of sound $c$, and impose the appropriate boundary conditions for the modes that are solutions to Eqs.\ (\ref{cde}). For more general profiles a numerical analysis can be performed, see for example \cite{uno}.

\subsection{Acoustic black holes and the Hawking effect}

A simple analytical model of an acoustic black hole \cite{Recati:2009ya} can be obtained by gluing two semi-infinite stationary and homogeneous 1D condensates, one subsonic ($x<0$) and the other supersonic ($x>0$), along a spatial discontinuity at $x=0$ (see \cite{Balbinot:2012xw}, to which we refer for more detailed explanations throughout this paragraph, and references therein): $c(x)=c_l\theta(-x) + c_r \theta(x)$. We take $v_0<0$, i.e. the flow is from right to left and $c_l<|v_0| < c_r$.
We denote the modes solutions in each homogeneous region and corresponding to the fields $\phi$ and $\varphi$ as
\bea
\phi_{\omega}=D(\omega)e^{-iwt+ik(\omega)x}\ ,\qquad \varphi_{\omega}=E(\omega)e^{-iwt+ik(\omega)x}\ ,
\eea
so that the equations (\ref{cde}) become
\bea\non\label{gupa}
\left[ (\omega-v_0k) - \frac{\xi c k^2}{2}  -\frac{c}{\xi} \right] D(\omega) &=& \frac{c}{\xi} E(\omega)\ , \\
\left[ - (\omega-v_0k) - \frac{\xi c k^2}{2} - \frac{c}{\xi}\right] E(\omega) &=& \frac{c}{\xi} D(\omega)  \ ,
\eea
while the normalization condition (\ref{nor}) gives
\bea\label{nodia}
 |D(\omega)|^2 - |E(\omega)|^2={1\over 2\pi  n_0}\Big|\frac{dk}{dw}\Big|\ .
\eea
The combination of the two Eqs. (\ref{gupa}) gives the Bogoliubov dispersion relation for a one-dimensional Bose liquid flowing at constant velocity
\begin{equation}\label{nrela}
(\omega-v_0k)^2=c^2\left(k^2+ \frac{\xi^2 k^4}{4}\right)
\end{equation}
containing the positive and negative norm branches $w-v_0k=\pm c\sqrt{k^2+\frac{\xi^2k^4}{4}}\equiv \pm \Omega(k)$  which, for the subsonic and supersonic regions, are given respectively in Figs. 1 and 2. 
\begin{figure}[h] \centering \includegraphics[angle=0, height=2in] {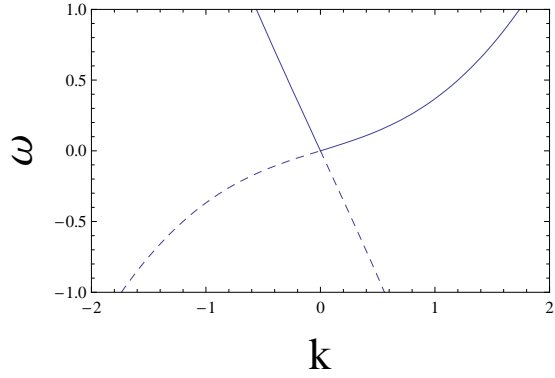}
\caption{Dispersion relation in the subsonic region ($w$ is given in units of the chemical potential and $k$ in units of the healing length).}
\end{figure}
\begin{figure}[h] \centering \includegraphics[angle=0, height=2in] {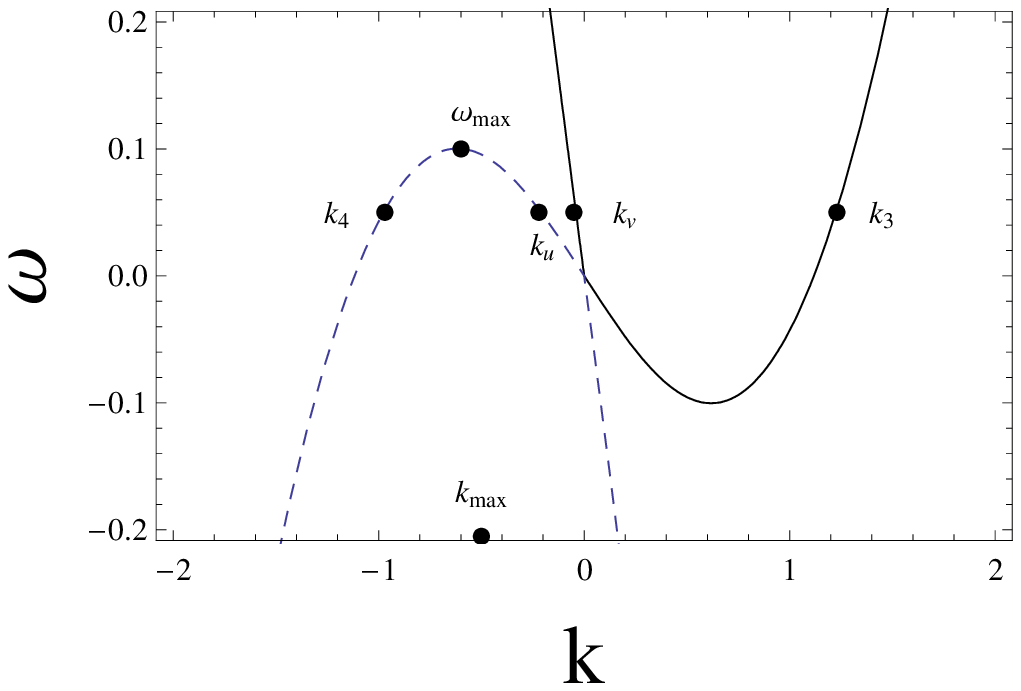}
\caption{Dispersion relation in the supersonic region (again, $w$ is given in units of the chemical potential and $k$ in units of the healing length).}
\end{figure}
Moreover, inserting the relation between $D$ and $E$ from (\ref{gupa}) into (\ref{nodia}) we find the mode normalizations
\bea\label{eq:normdispersion}
D(\omega) &=&  \frac{\omega -v k+\frac{c\xi k^2}{2}}{\sqrt{4\pi  n_0 c\xi k^2\left| (\omega-vk) \left(\frac{dk}{d\omega}\right)^{-1} \right| }},\nonumber\\
E(\omega) &=& -\frac{\omega -v k-\frac{c\xi k^2}{2}}{\sqrt{4\pi  n_0c\xi k^2\left| (\omega-vk) \left(\frac{dk}{d\omega}\right)^{-1} \right| }},
\eea
where $k=k(\omega)$ are the roots of the quartic equation (\ref{nrela}) at fixed $\omega$. 

In the subsonic case eq. (\ref{nrela}) admits two real and two complex solutions. Regarding the real solutions,  Fig. 1, we call $k_v \ (\sim \frac{\omega}{v_0-c} + O(\xi^2))$ and $k_u \ (\sim \frac{\omega}{v_0+c}+ O(\xi^2))$  the ones corresponding to, respectively, negative and positive group velocity $v_g={d\omega\over dk}$ (the other two complex conjugated solutions correspond to, respectively, a spatially decaying $k_d$ and growing $k_g$ modes). In the supersonic case, see Fig. 2, we see that for the most interesting regime ($\omega<\omega_{max}\sim \frac{1}{\xi}$), there are now four real solutions, corresponding to four propagating modes: $k_v, k_u$ (present also in the hydrodynamical limit $\xi=0$) and $k_3, k_4(\sim \frac{1}{\xi})$, two of which ($k_u$ and $k_4$) belong to the negative norm branch. 

To find modes evolution for all $x$ one needs to write down the general solutions for  $\phi$ ($\varphi$) in the left supersonic ($l$) and the right subsonic ($r$) regions (we restrict to the case $\omega<\omega_{max}$) 
\begin{eqnarray}
 \phi_{\omega}^{l} &=& e^{-i\omega t}\left[D_v^{l}A_v^{l}e^{ik_v^{l}x}+D_u^{l}A_u^{l}e^{ik_u^{l}x}+D_{3}^{l}A_3^{l}e^{ik_3^{l}x}+D_{4}^{l}A_4^{l}e^{ik_4^{l}x}\right], \nonumber\\ \phi_{\omega}^{r} &=& e^{-i\omega t}\left[D_v^{r}A_v^{r}e^{ik_v^{r}x}+D_u^{r}A_u^{r}e^{ik_u^{r}x}+d_{\phi}A_d^{r}e^{ik_d^{3}x}+G_{\phi}A_g^{r}e^{ik_g^{r}x}\right], 
  \end{eqnarray}
 (the expansions for $\varphi$ are the same up to the replacement $D\to E$) and 
   impose, from eqs.\ (\ref{cde}),  the matching conditions 
\begin{equation}\label{matchingaa}
[\phi]=0,\, [\phi']=0,\, [\varphi]=0,\, [\varphi']=0,
\end{equation}
where [ ] indicates the variation across the jump at $x=0$, allowing to write down the relations between left and right amplitudes $A$ through a scattering matrix $M_{scatt}$ in the form
\begin{equation}\label{scama}
     \left( \begin{array}{c}
       A_v^l \\
       A_u^l \\
       A_3^l \\
       A_4^l \\
     \end{array} \right)
   =M_{scatt} \left(
                \begin{array}{c}
                             A_v^r \\
       A_u^r \\
       A_d^r \\
       A_g^r \\
                \end{array}
              \right) \ .
\end{equation} 
This allows to construct explicitly the decomposition of the field operator $\hat \phi$ in terms of the  ``in'' and ``out'' basis. The ``in'' basis is constructed with $\phi^{in}$ modes propagating from the asymptotic regions ($x\to \pm\infty$) towards the discontinuity ($x=0$), while the 'out' basis is constructed with modes $\phi^{out}$  propagating away from the discontinuity to $x=\pm\infty$. Looking at Figs. 1 and 2, we see that unit amplitude modes defined on the left moving $k_v^r$ and right-moving $k_3^l, k_4^l \ (\sim \frac{1}{\xi})$ momenta define the ingoing scattering states, while unit amplitude modes defined on the right moving $k_u^r$ and left-moving $k_v^l,k_u^l$ momenta  define the outgoing scattering states. 
One can then write down the ``in'' decomposition in terms of the ``in'' scattering states
\begin{eqnarray}
\hat\phi&=&\int_{0}^{\omega_{max}}d\omega\left[\hat a_{\omega}^{v,in}\phi_{\omega}^{v,in}+\hat a_{\omega}^{3,in}\phi_\omega^{3,in}+\hat a_{\omega}^{4,in\dagger}\phi_\omega^{4,in}+h.c. \right]\ \ \ \ \ \ \ \ \
\end{eqnarray}
or, equivalently, on the basis of the ``out'' scattering ones. Note that since $k_4^l$ belongs to the negative norm branch the corresponding ``in'' mode $\phi^{in}_{4,l}$ is multiplied by a creation operator  $\hat a_{\omega}^{4,in\dagger}$ (the same thing happens, in the ``out'' decomposition, for $\phi^{out}_{u,l}$).
Using (\ref{scama}) one can construct the $3\times 3$ S-matrix relating  $\phi^{in}$ and $\phi^{out}$ modes  
\begin{eqnarray}\label{eq:outinaa}
  \phi_\omega^{v,in}&=& S_{vl,vr} \phi_\omega^{v,out}+S_{ur,vr} \phi_\omega^{ur,out}+S_{ul,vl} \phi_\omega^{ul,out}\ , \\
  \phi_\omega^{3,in} &=& S_{vl,3l} \phi_\omega^{v,out}+S_{ur,3l} \phi_\omega^{ur,out}+S_{ul,3l} \phi_\omega^{ul,out}\ , \\
  \phi_\omega^{4,in} &=& S_{vl,4l} \phi_\omega^{v,out}+S_{ur,4l} \phi_\omega^{ur,out}+S_{ul,4l} \phi_\omega^{ul,out}\ .
\end{eqnarray}
which is not trivial since it mixes positive and negative norm modes. As a consequence, the Bogoliubov transformation between `in'' and ``out'' creation and annihilation operators
 is also not trivial because it mixes creation and annihilation operators.
This has the crucial consequence that the  ``in" and  ``out" Hilbert spaces are not unitary related, in particular the corresponding vacua are different, i.e.  $|0,in\rangle \neq   |0,out \rangle$. The physical consequence is that if we prepare the system in the
$|0,in\rangle$ vacuum state, so there are no incoming phonons  at $t=-\infty$, we will have, at late times, outgoing quanta on both sides of the horizon: the vacuum has spontaneously emitted phonons, mainly in the $k_u^r$ channel (Hawking quanta) and $k_u^l$ (partners). The analytical calculations show that the number of emitted Hawking quanta \cite{Recati:2009ya} \begin{equation}\label{hq}
\langle 0,in|\hat a^{ur, out\ \dagger}_\omega  \hat a^{ur, out}_\omega|0,in\rangle =   |S_{ur,4l}|^2 \sim \frac{1}{\omega} \end{equation} and partners
\begin{equation}\label{pa}
\langle 0,in|\hat a^{ul, out\ \dagger}_\omega  \hat a^{ul, out}_\omega|0,in\rangle = |S_{ul,4l}|^2 \sim \frac{1}{\omega}
\end{equation}  follow an approximate (low-frequency) thermal $\frac{1}{w}$ spectrum \cite{cinque}, the proportionality factor allowing to identify a Hawking temperature ($\sim\frac{1}{\xi}$) in this idealised setting.
We can understand the mechanism by which Hawking radiation is emitted by looking, using (\ref{def}), at the equal-time density-density correlator 
\begin{equation}\label{densitya}
G^{(2)}(t;x,x')\equiv \frac{1}{n_0^2}\lim_{t\rightarrow t'}\langle 0,in| \hat n^1 (t,x), \hat n^1 (t',x') |0,in\rangle\ \ ,
\end{equation}
whose main contribution in the $x<0$ and $x'>0$ sector comes from the $ul-ur$ term 
\begin{equation}\label{stat}
\frac{1}{n_0^2}\langle \hat n_1 \hat n_1 \rangle |_{Hawking}  \sim Re \int_0^{w_{max}}dw S_{ul,4l}S_{ur,4l}^* (\phi_{u,l}^{w,out} + \varphi_{u,l}^{w,out})(\phi_{u,r}^{w,out*} + \varphi_{u,r}^{w,out*}) \sim 
\frac{\sin\left[\omega_{max}(\frac{x'}{v_0+c_r}-\frac{x}{v_0+c_l})\right]}{\frac{x'}{v_0+c_r}-\frac{x}{v_0+c_l}}\ .\end{equation}
The existence of the peak at \begin{equation}\label{peak}\frac{x'}{v_0+c_r}=\frac{x}{v_0+c_l}\end{equation}  was first pointed out in \cite{Balbinot:2007de} in the hydrodynamical approximation using QFT in curved space techniques. 
The physical picture that emerges is that Hawking quanta and partners are continuously created in pairs from the horizon at each time $t$, propagate on opposite directions at speeds
$v_0+c_l<0$ and $v_0+c_r>0$ and after time $\Delta t$ are located at $x$ and $x'$ related as in (\ref{peak}). The existence of the Hawking peak was nicely confirmed by numerical 'ab initio' simulations with more realistic configurations performed in  \cite{fnum}. How correlation measurements can reveal the quantum nature of Hawking radiation is discussed in \cite{due}.

\subsection{Analog dynamical Casimir effect}
A distinct type of pair-creation takes place in time-dependent backgrounds, one important example being quantum particle creation in cosmology. We will study the analogue of this phenomena in BECs
with another simple model in which the speed of sound  has a steplike discontinuity in time at t=0 separating two 'initial' and 'final' infinite homogeneous condensates: $c(t)=c_{in}\theta(-t)+c_{fin}\theta(t)$, see for more details \cite{ftemp}, \cite{Mayoral:2010un}.  The modes solutions in the initial ($t<0$) and final ($t>0$) regions are now of the type 
\bea
\phi_{k}=D(k)e^{-iw(k)t+ikx}\ ,\qquad\varphi_{k}=E(k)e^{-iw(k)t+ikx}\ ,
\eea
for which Eqs. (\ref{cde}) become
\begin{eqnarray}\label{eq:primerorden22aa}
\left[-(\omega-vk)+\frac{c\xi k^2}{2}+\frac{c}{\xi}\right]D(k)&=&-\frac{c}{\xi}E(k)\ , \nonumber\\
\left[(\omega-vk)+\frac{c\xi k^2}{2}+\frac{c}{\xi}\right]E(k)&=&-\frac{c}{\xi}D(k)\ ,
\end{eqnarray}
while the normalization condition (\ref{nor}) yields
\begin{equation}\label{nodi}
 |D(k)|^2 - |E(k)|^2=\frac{1}{2\pi  n_0}
 \end{equation}
 giving
\begin{eqnarray}\label{eq:normdispersionk}
   &&D(k) =  \frac{\omega -v k+\frac{c\xi k^2}{2}}{\sqrt{4\pi  n_0 c\xi k^2\left| (\omega-vk) \right| }}\ ,\nonumber\\
  &&E(k) = -\frac{\omega -v k-\frac{c\xi k^2}{2}}{\sqrt{4\pi  n_0 c\xi k^2\left| (\omega-vk)  \right| }}\ .
\end{eqnarray}
Here, $\omega=\omega(k)$ corresponds to the two real solutions to Eq.\  (\ref{nrela}), which is quadratic in $\omega$ at fixed $k$.  These read
\begin{eqnarray}
\omega_+(k)&=&vk+\sqrt{c^2k^2+\frac{c^2k^4\xi^2}{4}}\equiv vk -\Omega(k)\ ,\nonumber\\
\omega_-(k)&=&vk-\sqrt{c^2k^2+\frac{c^2k^4\xi^2}{4}}\equiv vk -\Omega(k)\ ,
\end{eqnarray}
where $\omega_+(k)$ corresponds to the positive norm branch, and $\omega_-(k)$ to the negative norm one. 
To find modes evolution for all $t$ one first write down the general solutions in the initial and final regions for $\phi$ ($\varphi$)
\be
  \phi_{k}^{fin(in)}=e^{ik x}\left[D_{fin(in)}^+(k)A_{fin(in)}e^{-i\omega_+^{fin(in)}(k)t}+D_{fin(in)}^-(k)B_{fin(in)}e^{-i\omega_-^{fin(in)}(k)t}\right] \ee
(for $\varphi$ we have the same expansion with $D$ replaced by $E$) and impose the matching conditions, from (\ref{cde}),  
\begin{equation}\label{matching}
[\phi]=0, \,[\varphi]=0\ ,
\end{equation}
where [ ] now indicates the variation across the discontinuity at $t=0$. 
They allow the final amplitudes $A^{fin}$ to be related to the initial ones $A^{in}$ through the matrix $M_{Bog}$ 
\begin{equation}
\label{eq:matchingtc}
     \left( \begin{array}{c}
       A_{fin} \\
       B_{fin} \\
     \end{array} \right)
   =M_{bog} \left(
                \begin{array}{c}
                  A_{in} \\
                  B_{in} \\
                \end{array}
              \right) 
\end{equation}
where 
 \begin{equation}
\label{eq:bogoliubovda}
M_{bog}=\frac{1}{2\sqrt{\Omega_{in}\Omega_{out}}}\left(
            \begin{array}{cc}
               \Omega_{in}+\Omega_{fin} &
               \Omega_{fin}-\Omega_{in} \\
               \Omega_{fin}-\Omega_{in} &
               \Omega_{in}+\Omega_{fin} \\
            \end{array}
          \right) \ .
\end{equation}  
One can easily construct 'in' and 'fin' decompositions for the field $\hat\phi$ by considering initial (final) unit amplitude positive norm modes ($A_{in(fin)}=1,\ B_{in}=0$) $\phi_k^{in(fin)}$ ($\varphi_k^{in(fin)}$) modes for all $t$ using (\ref{eq:matchingtc})
\begin{equation}\label{phia}
\hat \phi(t,x)^{in(fin)} = \int_{-\infty}^{\infty} dk\left[\hat a_{k}^{in(fin)}\phi_k^{in(fin)}+a_{k}^{in(out)\dagger}\varphi_k^{in(fin)*}\right]\ .
\end{equation} 
The modes are related through a nontrivial Bogoliubov transformation mixing positive and negative norm modes \begin{equation}
\phi_k^{in}=\alpha_k\phi^{fin}+\beta_k \phi_{-k}^{out*},
\end{equation}
where 
\begin{equation}\label{bove}
 \alpha_k=\frac{\Omega_{in}+\Omega_{fin}}{2\sqrt{\Omega_{in}\Omega_{fin}}}\ ,\qquad  \beta_k=\frac{\Omega_{fin}-\Omega_{in}}{2\sqrt{\Omega_{in}\Omega_{fin}}}\ \end{equation}
and, consequently, also the relation between ''in'' and ''fin'' annihilation and creation operators will mix annihilation and creation operators, implying again that the two decompositions are inequivalent and in particular the two vacuum states $|0,in\rangle$ and $|0,fin\rangle$ are different.  The physical consequence is that the steplike discontinuity at $t=0$ will induce particle creation, the features of which can be understood by looking at the time-dependent terms of the one-time density-density correlator which in the hydrodynamical limit reads
\begin{equation}\label{gtwoto}
\begin{array}{cc}
G^{(2)}(t,x,x')|_{dyncas}\sim 
 \frac{1}{(2c_{fin}t -(x-x'))^2} + \frac{1}{(2c_{fin}t -(x'-x))^2}\ .
\end{array}
\end{equation}
At $t=0$ and everywhere in space correlated pairs of particles with opposite momentum are created out of the vacuum state,
with velocities $v-c_{fin}$ (left-moving) and $v+c_{fin}$ (right-moving). At time $t$ such particles are separated by a distance \begin{equation}\label{pedy} |x-x'|=2c_{fin}t\ ,\end{equation} which is indeed the correlation displayed in (\ref{gtwoto}).  This effect was recently observed in \cite{Jaskula:2012ab}  by considering homogeneous condensates with trapping potential $V_{ext}$ rapidly varying in time, where correlation functions  in velocity/momentum space were measured.

\subsection{Amplification  of the Hawking signal in density correlators}
It has been argued by Cornell \cite{cornell} that a way to amplify the Hawking signal in density-density correlators is to 
reduce the interactions shortly before measuring the density correlations. Since  $c=\sqrt{\frac{gn_0}{m}}$, reducing $g$ means that the speed of sound is also reduced.
 We will model this situation by matching our idealised acoustic black hole configuration  of section 3.1  with a final infinite homogeneous condensate characterised by a 
small sound velocity $c_{fin} $ ($< c_l < c_r$), see Fig. 3. To study this situation, in which a spatial step-like discontinuity in $c$ at $x=0$ is combined with a temporal step-like discontinuity at some $t=t_0$, we shall use the tools introduced in the previous two subsections. 
\begin{figure}[h] \centering \includegraphics[angle=0, height=2in] {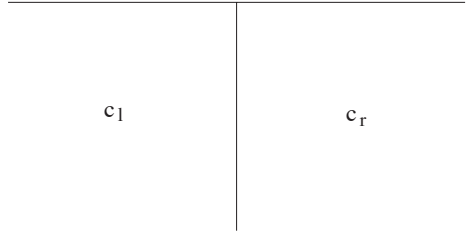}
\caption{Spacetime diagram sketch of a spatial step-like discontinuity (acoustic black hole-like) followed by a temporal one leading to a final homogeneous configuration.}
\end{figure}
We shall calculate the density-density correlator in the 'in' vacuum by expanding the density operator in the 'in' decomposition 
\begin{equation}
\hat n^1(t,x) \simeq n_0 \int_0^{\omega_{max}}dw [ \hat a_\omega^{v,in} (\phi_{v,r}^{in} + \varphi_{v,r}^{v,in}) + \hat a_\omega^{3,in} (\phi_{3,l}^{in} + \varphi_{3,l}^{in}) + \hat a_\omega^{4,in \dagger} (\phi_{4,l}^{in} + \varphi_{4,l}^{in}) + h.c.]
\end{equation}
to get
\begin{equation}
\frac{1}{n_0^2}\langle 0,in| \hat n_1 \hat n_1 |0,in\rangle = \int_0^{w_{max}}dw [ (\phi_{v,r}^{in} + \varphi_{v,r}^{v,in})(\phi_{v,r}^{in*} + \varphi_{v,r}^{v,in*})+(\phi_{3,l}^{in} + \varphi_{3,l}^{in})(\phi_{3,l}^{in*} + \varphi_{3,l}^{in*})+(\phi_{4,l}^{in*} + \varphi_{4,l}^{in*})(\phi_{4,l}^{in} + \varphi_{4,l}^{in})]\ .
\end{equation}
By expressing the 'in' modes in terms of the 'out' modes and in the absence of the temporal step-like discontinuity (say, $t_0\to +\infty$) the Hawking signal is given by
\begin{equation}\label{stat}
\frac{1}{n_0^2}\langle 0,in|\hat n_1 \hat n_1 |0,in \rangle  |_{Hawking} \sim Re \int_0^{w_{max}}dw S_{ul,4l}S_{ur,4l}^* (\phi_{u,l}^{w,out} + \varphi_{u,l}^{w,out})(\phi_{u,r}^{w,out*} + \varphi_{u,r}^{w,out*})\ .
\end{equation}
In the presence of the temporal step-like discontinuity we need to evolve the relevant modes $\phi_{u,l}^{out}$ ($\varphi_{u,l}^{out}$) and $\phi_{u,r}^{out}$ ($\varphi_{u,r}^{out}$), at the same value of $w$, across the discontinuity
at $t=t_0$. Going to the $k$ basis and considering $k, k'$ small ($w\sim  (v+c_l)k \sim (v+c_r)k'$) we have
\begin{equation}\label{lfink}
 \phi_{u,l}^{k,out}+ \varphi_{u,l}^{k,out}\sim  \sqrt{\frac{|k|}{c_l}}e^{-i(v+c_l)kt+ikx}\to  \sqrt{\frac{|k|}{c_{fin}}}e^{ikx}(\alpha e^{-i(v+c_{fin})kt}+\beta e^{-i(v-c_{fin})kt})\ ,
 \end{equation}
 where $\alpha=\frac{c_l+c_{fin}}{2\sqrt{c_lc_{fin}}}e^{i(c_{fin}-c_l)kt_0},\ \beta=\frac{c_{fin}-c_l}{2\sqrt{c_lc_{fin}}}e^{-i(c_{fin}+c_l)kt_0}$ and
 \begin{equation}\label{rfink}
 \phi_{u,r}^{k',out}+ \varphi_{u,r}^{k',out}\sim  \sqrt{\frac{k'}{c_r}}e^{-i(v+c_r)k't+ik'x'}\to  \sqrt{\frac{k'}{c_{fin}}}e^{ik'x'}(\alpha' e^{-i(v+c_{fin})k't}+\beta' e^{-i(v-c_{fin})k't})\ ,
 \end{equation}
 with $\alpha'=\frac{c_r+c_{fin}}{2\sqrt{c_rc_{fin}}}e^{i(c_{fin}-c_r)k't_0},\ \beta'=\frac{c_{fin}-c_r}{2\sqrt{c_rc_{fin}}}e^{-i(c_{fin}+c_r)k't_0}$. 
 
 It is useful to rewrite (\ref{lfink}) and (\ref{rfink}) in terms of $w$ to compare with the standard result without the temporal step-like discontinuity
 \begin{equation}\label{lfinw}
 \phi_{u,l}^{w,out}+ \varphi_{u,l}^{w,out}\sim  \sqrt{\frac{w}{c_l}}e^{-iwt+i\frac{wx}{v+c_l}}\to  \sqrt{\frac{w}{c_{fin}}}e^{i\frac{wx}{v+c_l}}\left(\alpha e^{-i\frac{(v+c_{fin})}{v+c_l}wt}+\beta e^{-i\frac{(v-c_{fin})}{v+c_l}wt}\right)\ ,
 \end{equation} 
  \begin{equation}
 \phi_{u,r}^{w,out}+ \varphi_{u,r}^{w,out}\sim  \sqrt{\frac{w}{c_r}}e^{-iwt+i\frac{wx'}{v+c_r}}\to  \sqrt{\frac{w}{c_{fin}}}e^{i\frac{wx'}{v+c_r}}\left(\alpha' e^{-i\frac{(v+c_{fin})}{v+c_r}wt}+\beta' e^{-i\frac{(v-c_{fin})}{v+c_r}wt}\right). \label{rfinw} \ \ \ \ \ 
 \end{equation} 
 The standard result is, from (\ref{stat}),  a stationary peak at \begin{equation}\label{statpeak} \frac{x}{v+c_l}-\frac{x'}{v+c_r}=0 \end{equation}  weighted 
 by the $w$-independent factor (see (\ref{hq},\ref{pa})) \begin{equation} \label{statsign}\frac{w}{\sqrt{c_lc_r}}S_{ul,4l}S_{ur,4l}^*\ . \end{equation} The effect of the temporal step-like discontinuity at $t=t_0$, see (\ref{lfinw}, \ref{rfinw}), is to modify this signal
 into a main signal located at 
 \begin{equation}\label{nono} \frac{x}{v+c_l}-\frac{x'}{v+c_r}=\left(\frac{(v+c_{fin})}{v+c_l}-\frac{(v+c_{fin})}{v+c_r}\right)(t-t_0) \end{equation}  
with strength
\begin{equation}\label{sisi} \frac{w\alpha\alpha'}{c_{fin}}S_{ul,4l}S_{ur,4l}^* \  \end{equation}
and three smaller signals located at $\frac{x}{v+c_l}-\frac{x'}{v+c_r}=(\frac{(v+c_{fin})}{v+c_l}-\frac{(v-c_{fin})}{v+c_r})(t-t_0)$, $\frac{x}{v+c_l}-\frac{x'}{v+c_r}=(\frac{(v-c_{fin})}{v+c_l}-\frac{(v+c_{fin})}{v+c_r})(t-t_0)$, $\frac{x}{v+c_l}-\frac{x'}{v+c_r}=(\frac{(v-c_{fin})}{v+c_l}-\frac{(v-c_{fin})}{v+c_r})(t-t_0)$   with strengths given by (\ref{sisi}) in which $\alpha\alpha'$ is substituted, respectively, by $\alpha\beta'$, $\alpha'\beta$ and $\beta\beta'$. 

We see immediately that we loose the stationarity of the Hawking signal (\ref{statpeak}) and that the main signal is multiplied by \begin{equation}\label{amp}  \eta=\frac{\sqrt{c_rc_l}}{c_{fin}}\alpha\alpha'=
\frac{(c_{fin}+c_l)(c_{fin}+c_r)}{4c_{fin}^2} \end{equation}
with respect to the standard result. This results indeed in an amplification (i.e. the above term is $>1$) when $c_{fin}<c_l,c_r$. Being the Hawking peak expected to be of order $5\times 10^{-3}$ for realistic experimental settings \cite{fnum} (where $c_r=2c_l$ was considered), we obtain an amplification factor $\eta=\frac{15}{4}$ for $c_r=2c_l=4c_{fin}$.

 \section{Conclusions}

In this paper we have briefly reviewed the analysis of the analog Hawking effect and of the analog dynamical Casimir effect  by considering 
simple analytical models of Bose-Einstein condensates in which the speed of sound has step-like discontinuities. We focussed in the study of the density-density
correlators which show, in the former case, the existence of a characteristic stationary Hawking quanta - partner peak located at (\ref{peak}) and, in the latter, of a time dependent feature (\ref{pedy}).
Following a suggestion by Cornell, we combined these two analysis to construct a model in which the atoms' interactions are rapidly lowered (and so the speed of sound) before the correlations are measured. 
This results in an amplification of the main Hawking peak, now time-dependent and located at (\ref{nono}), by the factor (\ref{amp}) that could be useful in the experimental search.

\acknowledgments
We thank I. Carusotto for useful discussions.

{}
\end{document}